\documentclass[12pt]{article}
\usepackage{amssymb}
\usepackage[dvips]{graphics}
\usepackage{epsfig}
\newcommand{\be}{\begin{equation}}
\newcommand{\ee}{\end{equation}}

\begin{document}
\title{\LARGE{\textbf{The packing of two species of polygons \\ on the square lattice}}}
\author{\small{David Dei Cont and Bernard Nienhuis}}
\date{\normalsize \emph{Instituut voor Theoretische Fysica, Universiteit van Amsterdam,\\
Valckenierstraat 65, 1018 XE Amsterdam, The Netherlands} \\
deicont@science.uva.nl, nienhuis@science.uva.nl
}
\maketitle
\thispagestyle{empty}
\begin{abstract}
We decorate the square lattice with two species of polygons under the
constraint that every lattice edge is covered by only one polygon and 
every vertex is visited by both types of polygons.
We end up with a 24 vertex model which is known in the literature as
the fully packed double loop model ($\textrm{FPL}^2$).
In the particular case in which the fugacities of the polygons are
the same, the model admits an exact solution.
The solution is obtained using coordinate Bethe ansatz and provides
a closed expression for the free energy.
In particular we find the free energy of the four colorings model
and the double Hamiltonian walk and recover the known entropy of 
the Ice model.
When both fugacities are set equal to two the model 
undergoes an infinite order phase transition.
\\
\end{abstract}
\clearpage
\section{Introduction}
Loop models appear in a variety of contexts:
as diagrammatic expansion of the Potts and
the O(n) models \cite{D1}, and as models to
describe polymers in several phases.
The conformal properties were investigated
using the Coulomb gas mapping
\cite{BERNARD3}.
An exact expression for the free energy, 
in the thermodynamic limit, 
was found using the coordinate Bethe ansatz method
and solving linear integral equations
\cite{B3,B4}.
Since then, new loop models have been studied
on the square lattice \cite{BA3,JK1},
on decorated lattices \cite{J4,HIGUCHI1},
on a random lattice \cite{KOSTOV},
and in three dimensions \cite{HIGUCHI2}.
The resolution techniques have been improved
both on the 
numerical side, 
introducing the connectivity basis
\cite{BLOTE2,BLOTE3,JK1}
and on the analytical front
adding the seam
\cite{BA1,BA1bis},
making use of the algebraic Bethe ansatz
\cite{R},
mapping to an interface model
\cite{KONDEV1,JK1}
and exploiting conformal field theory
\cite{DOTSENKO}.

Here we are interested in 
the fully packed double loop model 
($\textrm{FPL}^{2}$)
on the square lattice.
It is defined by filling the square lattice
with loops in two colors, drawn along the lattice edges
in such a way that every bond is covered by one
loop and every site is visited by both types of loops.
It was introduced for the first time in
\cite{KH}
as an alternative representation of 
the four colorings model on the square lattice.
It is a natural generalization of the
fully packed one loop model on the honeycomb lattice
\cite{BLOTE1,BA2,R,KONDEV2},
which is the loop representation of the 
three colorings model on the hexagonal lattice
\cite{B1}.

Jacobsen and Kondev mapped the $\textrm{FPL}^{2}$ model
onto a height model and postulate that the long wavelength
behavior is correctly described by a Liouville field theory
\cite{KONDEV4}.
This provides a geometrical view of conformal invariance
in two dimensional critical phenomena and a method to
access the critical properties of loop models exactly.
In particular, 
they succeed in calculating the central charge
and the critical exponents \cite{JK1,KONDEV3}. 

The $\textrm{FPL}^{2}$ model 
exhibits a rich phase diagram and provides a 
representation for previously studied models:
Ice model \cite{L1},
four colorings model \cite{KH},
dimer loop model \cite{RAGHAVAN},
double Hamiltonian walk \cite{KONDEV4},
compact polymers \cite{BA4}.
A generalization of the 
$\textrm{FPL}^{2}$ model,
which offers a unifying picture of the compact,
dense and dilute phases of polymers,
was proposed in \cite{J3} by relaxing the full packing constraint.
Another possible generalization corresponds to the
Flory model of polymer melting \cite{J2}.
Recently the $\textrm{FPL}^{2}$ model has been coupled
to two dimensional quantum gravity in order to study meanders
\cite{DIFRANCESCO}.

Here we find an exact expression for the free
energy, in the thermodynamic limit, 
when the loop fugacities are the same.
The solution relies on coordinate Bethe ansatz.
Unfortunately the compact polymer does not
belong to the solvable line.
The configurational
entropy for the double Hamiltonian walk \cite{KONDEV4,JK1}
and the four colorings model \cite{KH} is calculated.
We also confirm a conjecture 
on the average length of loops 
\cite{J1}.

The present work is organized as follows.
In section 2 we map the $\textrm{FPL}^{2}$
model into a 24 vertex model on the square lattice.
The transfer matrix is set up in section 3 and
diagonalized by means of coordinate Bethe ansatz (BA) in section 4.
The free energy, in the thermodynamic limit,
is calculated in section 5.
\section{The model}
\begin{figure}[!h]
\centerline{\epsfxsize=16cm \epsfbox{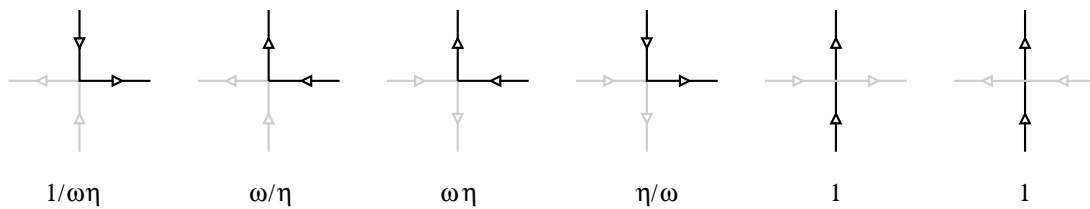}}
\caption{The 6 basic types of vertices of the 24 vertex model and the corresponding weights.
When $\omega = \eta$ the model admits an exact solution.
 \label{fig1}}
\end{figure}

We study a 24 vertex model on the square lattice.
Every edge is decorated with one of two types of arrows, black or grey ones.
The admitted configurations of arrows do not contain sources and sinks, so
that for both colors there is one arrow going in and one going out of each vertex.
The 24 possible vertices may be divided into 6 basic types depicted
in Fig.~\ref{fig1}, the others being related by $\pi/2$ rotations.
We can interpret each vertex as a couple
of lines turning clockwise or anticlockwise.
In this way the vertex model is mapped into an oriented loop model.
We are interested in the partition function of the unoriented loop model defined as:
\be
Z=\sum n_{b}^{N_{b}} n_{g}^{N_{g}} \label{FPLpartfunc}
\ee
where $N_{b}$ and $N_{g}$ are the respective numbers of black and grey loops.
We assign the fugacity $n_{b}$ for black loops and $n_{g}$ for grey loops.
The sum is over all the allowed loop configurations.
The fugacities of the loops are related to the weights of the vertex model by:
\be
\label{relation1}
\omega \equiv \textrm{exp}(\textrm{i} \, \pi \, e_{b}/4)
\qquad
\eta \equiv \textrm{exp}(\textrm{i} \, \pi \, e_{g}/4)
\qquad
n_{b}=2 \, \cos \pi \, e_{b}
\qquad
n_{g}=2 \, \cos \pi \, e_{g}
\ee
This is the two-flavor fully packed loop model ($\textrm{FPL}^{2}$)
on the square lattice investigated by Jacobsen and Kondev \cite{JK1}.
We will see that when the vertex weights are the same $\omega=\eta$,
the partition function of $\textrm{FPL}^2$ (\ref{FPLpartfunc})
can be computed exactly, in the thermodynamic limit,
making a coordinate Bethe ansatz. 
For particular values of the fugacity we recover
the partition function of previously studied models.
In particular $(n_{b},n_{g})=(2,2)$ corresponds
to the four colorings model on the square lattice \cite{KH}.
For $(n_{b},n_{g})=(1,1)$ we get the Ice model solved by Lieb \cite{L1}.
The single Hamiltonian walk is obtained for $(n_{b},n_{g})=(0,1)$ \cite{JK1}. 
Finally $(n_{b},n_{g})=(0,0)$ is the double Hamiltonian walk for which the
entropy has been correctly conjectured in \cite{JK1}, as we will show.
%
%
\section{The transfer matrix}
%
Here we set up a transfer matrix for the model following \cite{L1,B1}.
Consider a square lattice on a cylinder, made up by $L M$ vertices
and connect the vertices by vertical and horizontal bonds.
Starting from the bottom we have a circular row of L vertical bonds
followed by a row of L horizontal bonds and so on alternately.
Denote by $|\Psi\rangle$ the possible configuration of a row of L vertical
bonds, that is to say a precise assignment of arrows on each of the vertical
bonds. There are $4^L$ possible choices.
Let $|\Psi\rangle$ and $|\Psi^{\prime}\rangle$ be the states of two consecutive
rows of vertical bonds.
The intervening row of horizontal edges is decorated so that the L resulting vertices
are of the types shown in Fig.~\ref{fig1}.
To each allowed configuration of vertices we assign a weight given by the product
of the weight of each single vertex.
Then we sum over all the possible configurations of horizontal bonds compatible with
the vertical bonds and obtain the total weight denoted by 
$\langle \Psi^{\prime}| \mathbf{T} |\Psi\rangle$.
In this way we build up the transfer matrix $\mathbf{T}$ for the model.
The physics of the model is encoded in the eigenvalues
and the eigenvectors of the transfer matrix.
If periodic boundary conditions are imposed in the vertical direction, 
the partition function is simply the trace of the transfer matrix
raised to the power $M$: 
\be
Z=\textrm{Tr} \, \mathbf{T}^{M}
\ee
In the thermodynamic limit the free energy $f_{\infty}$
follows from the largest eigenvalue $\Lambda(L)$ of 
the $4^{L}$-square matrix $\mathbf{T}$ by:
\be
f_{\infty}=\lim_{L \rightarrow \infty} \frac{1}{L} \log \Lambda(L)
\ee
If we interpret the vertical axis of the cylinder as time 
(running downward, for convenience)
and the horizontal extend as space we end up with 
the time evolution of a one dimensional system.

In the following we are going to study the possible evolution of a generic state.
Impose periodic boundary conditions on the horizontal direction.
It is convenient to start from the state which consists of only gray arrows
pointing up (pseudo-vacuum).
This can match only with a successive row of gray arrows up.
There are two possibilities for the intervening row of horizontal bonds:
they must be black arrows, either running all to the right or all to the left.
The total weight is two.
The next step is to replace some of the gray arrows up by black arrows pointing up.
We call a black arrow up an \emph{ordinary particle}.
In this case the upper row can evolve in two different lower rows as illustrated
by Fig.~\ref{fig3}: all the particles move either one step to the right (Fig.~\ref{fig3}~a)
or one step to the left (Fig.~\ref{fig3}~b).
The product of the weight of each single vertex equals one, because
vertices with reciprocal weight come in pairs,
so that the total weight of the transition is set to one.
More complex patterns arise allowing for black arrows pointing down.
We refer to a black arrow down as a \emph{special particle}.
Now, besides the global shift, 
there are other possible evolutions of which an example is depicted in Fig.~\ref{fig3}~c.
The special particle can be connected via a half-loop
to the left or right neighboring ordinary particle.
Then in order to fulfill the boundary constraint 
a black half-loop has to be inserted in the lower row.
This half-loop can turn either clockwise or anticlockwise and must be located between 
any two particles other than the two already connected via the upper row.
Thus we end up with one half-loop in the upper row and one half-loop in the lower row.
Particles shift one step to the right or to the left according to their position
relative to the half-loops.
The total weight is completely determined by the orientation of the two half-loops.
A clockwise half-loop contributes $\omega^{2}$ and an anticlockwise one $\omega^{-2}$.
To complete the list of allowed configurations we have to examine the states
containing grey arrows down.
We will interpret the grey arrow down as a bound state between 
two black arrows 
(see Fig.~\ref{fig4}~b and Fig.~\ref{fig5}).
We leave the details to the next section.

A fundamental ingredient,
in seeking for an exact solution,
is the existence of conserved quantities.
Besides the conservation of black and grey
flux, there is an additional conservation law,
which is less evident.
We saw that at every application of the transfer
matrix, both ordinary and special particles,
shift one step.
Therefore the lattice can be divided into two sublattices
such that the number of particles on each sublattice remains
constant from row to row.
As a result the transfer matrix splits in a series of diagonal
blocks relating states in the same sector.
Each sector is completely fixed by specifying the number
of odd, even and special particles.
We are going to diagonalize this matrix making an 
appropriate ansatz on the structure of the eigenvectors.

Before concluding this section we make some remarks concerning
the relation between the loop and the 24 vertex model.
We assign the weights
$\textrm{exp}(\pm \textrm{i} \, \pi \, e_{b,g})$
to each oriented loop.
The orientation fixes the sign and the color selects
the phase $e_{b}$ or $e_{g}$ .
The partition sum of the oriented 
$\textrm{FPL}^{2}$
model may be cast in the form:
\be
Z
=
\sum
[\textrm{exp}(\textrm{i} \, \pi \, e_{b}) + \textrm{exp}(-\textrm{i} \, \pi \, e_{b})]^{N_{b}}
[\textrm{exp}(\textrm{i} \, \pi \, e_{g}) + \textrm{exp}(-\textrm{i} \, \pi \, e_{g})]^{N_{g}}
\ee
where the sum runs over all the allowed 
unoriented loop configurations.
Relating the unoriented loop fugacities to the oriented loop phases by
$n_{b,g}=2 \, \cos \pi \, e_{b,g}$,
we recover the partition function of the unoriented loop model 
(\ref{FPLpartfunc}).
The advantage of splitting the loop fugacity among the two possible orientations,
is that in this way the loop fugacity can be distributed along all the vertices
visited by the loop.
This is achieved assigning the weight 
$\textrm{exp}(\pm \textrm{i} \, \pi \, e_{b,g}/4)$
to every vertex visited by the oriented loop,
with sign $+$ for a clockwise turn and $-$ for the anticlockwise one.
Since for every oriented loop, on the square lattice, 
the difference between clockwise and anticlockwise turns is $\pm 4$,
multiplying over all the local vertex weights we recover the
correct global loop weight.
This should clarify relations (\ref{relation1})
and the choices made for the vertex weights in Fig.~\ref{fig1}.
It is important to notice that the previous argument does not
work for a loop winding round the cylinder because
in that case the difference between clockwise and anticlockwise
turns is zero.
In order to give the correct weight to non-contractable loops a seam 
\cite{BA1,BA1bis,BA2} is placed between two generic vertices.
This is well depicted in Fig.~\ref{fig3}.
The total weight of a generic transition will
factorize in the weight of each single vertex
and an extra factor which accounts of the seam.
The additional factor is set equal to $a$ ($a^{-1}$) 
when the seam cuts a right (left) pointing arrow.
For convenience we introduce a phase $\alpha$ setting
$a=\textrm{exp}(\textrm{i} \, \pi \, \alpha)$. 
In terms of the phases the partition function of the
unoriented $\textrm{FPL}^2$ model on the cylinder reads:
\be
Z
=
\sum
(2 \cos \pi \, e_{b})^{N_{b}}
(2 \cos \pi \, e_{g})^{N_{g}}
(2 \cos \pi \, \alpha)^{N_{a}}
\ee
where $N_{a}$ is the total (black+gray) 
number of uncontractable loops.
Notice that, in order to simplify the calculation,
we do not distinguish between gray and black
uncontractable loops.
%
\section{The Bethe ansatz}
%
%
Since the grey arrow down serves as bound state between two black arrows,
we restrict, for the moment,  
to the sector containing only ordinary and special particles.
Denote the generic state of a row of $L$ vertical bonds by 
$|\mathbf{x},\mathbf{r} \rangle$.
The vector $\mathbf{x}=(x_{1},\ldots,x_{N})$ 
gives the position of all the $N$ particles
while the vector $\mathbf{r}=(r_{1},\ldots,r_{m})$ 
specifies that those at position
$x_{r_{1}},\ldots,x_{r_{m}}$ are special ones.
The eigenvalue equation for the transfer matrix reads:
\be
\mathbf{T} \, \sum_{|\mathbf{x},\mathbf{r} \rangle} \Psi(\mathbf{x},\mathbf{r}) \,  |\mathbf{x},\mathbf{r} \rangle
=
\Lambda
\,
\sum_{| \mathbf{x},\mathbf{r} \rangle } \Psi(\mathbf{x},\mathbf{r}) \,  | \mathbf{x},\mathbf{r} \rangle
\ee
where the eigenstate of eigenvalue $\Lambda$
is a linear combination,
with coefficients $\Psi(\mathbf{x},\mathbf{r})$, 
of the base $|\mathbf{x},\mathbf{r}\rangle$.
In the coefficients the eigenvalue equation assumes the form:
\be
\label{CEE}
\sum_{|\mathbf{x}',\mathbf{r}'\rangle}
\langle \mathbf{x},\mathbf{r}|\mathbf{T}|\mathbf{x}',\mathbf{r}'\rangle
\Psi(\mathbf{x}',\mathbf{r}')=
\Lambda \Psi(\mathbf{x},\mathbf{r})
\ee
Here we examine the sector with $N$ particles in total,
($N-1$ ordinary + $1$ special), located at position
$x_{1}, \ldots, x_{N}$, where $x_{i+1}-x_{i}>2$,
and suppose that the $j$th particle is a special one.
Moreover, we assume that the seam is placed between
two ordinary particles.
In order to solve the eigenvalue problem we make the following Ansatz
on the form of the wave function:
\be
\label{BetheAnsatz}
\Psi_{j}(x_{1},\ldots,x_{N}) 
\equiv  
\sum_{\mathbf{p}} A_{p_{1}\ldots p_{N}} \psi_{j}(\mathbf{p}) \, \prod_{i=1}^{N} z_{p_{i}}^{x_{i}}
\ee
where the sum runs over all permutations 
$\mathbf{p}=(p_{1},\ldots,p_{N})$
of the numbers 
$1,\ldots,N$.
The goal is to chose 
$A_{\mathbf{p}}$, $\psi_{j}(\mathbf{p})$
and the complex numbers 
$z_{p_{1}},\ldots,z_{p_{N}}$ so that (\ref{CEE})
is satisfied.
At this stage the structure of the coefficient
$\psi_{j}(\mathbf{p})$,
the behavior of $A_{p_{1} \ldots p_{N}}$ under the
exchange of two consecutive indices,
and the dependence of the
eigenvalue $\Lambda$ on 
$z_{p_{1}},\ldots,z_{p_{N}}$,
is still unknown.
In the following we are going to determine it.

\begin{figure}[!t]
\centerline{\epsfxsize=15cm \epsfbox{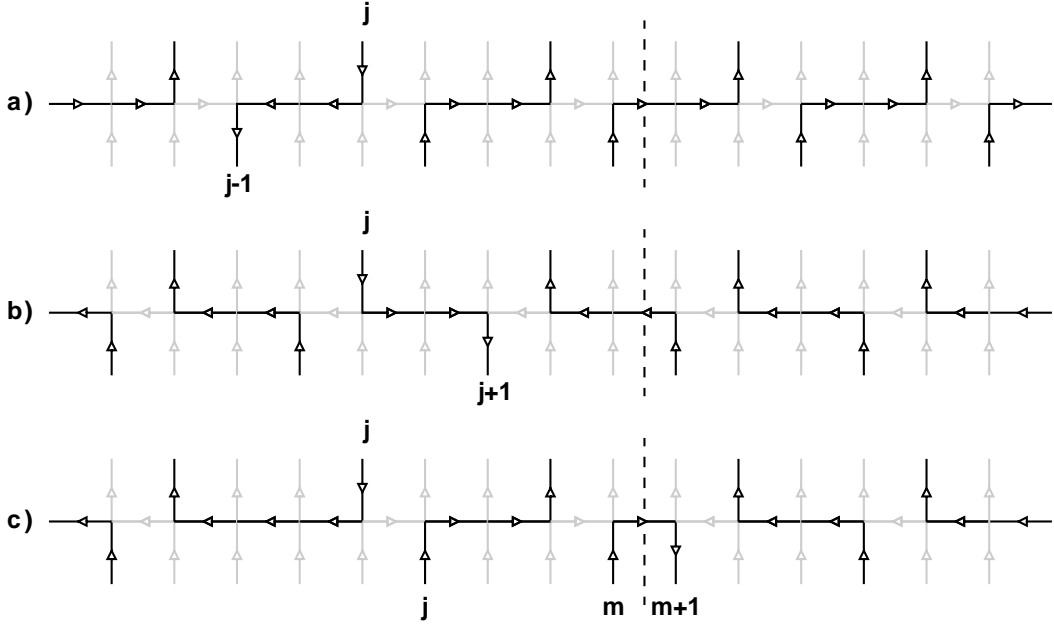}}
\caption{Four ordinary particles (black arrows up)
and one special particle (black arrow down).
Dashed line indicates the position of the seam.
The special particle, in the upper row, 
is the second one from the left ($j=2$). 
The possible evolutions of the upper row are:
a) Shift to the right.
b) Shift to the left. 
c) Appearance of the black half-loops.
In the lower row the special particle is the fourth one ($m+1=4$).
} \label{fig3}
\end{figure}  

Inserting the ansatz (\ref{BetheAnsatz}) 
into the eigenvalue equation (\ref{CEE}),
we see that a sufficient condition for the 
validity of (\ref{CEE}) is that the
coefficient $\psi_{j}(\mathbf{p})$ obeys, for any
permutation $\mathbf{p}$, the following 
evolution equation:
\begin{eqnarray}
\label{evolutionequation}
\Lambda \psi_{j}
& = &
a \, z_{1:N} \, \psi_{j-1} + \frac{1}{a \, z_{1:N}} \, \psi_{j+1}
\nonumber
\\ 
& + &
\omega^{2}
\sum_{m=j}^{j+N-2} 
(\omega^{-2} \, \alpha_{m} \, \psi_{m}+ \omega^2 \, \beta_{m+1} \, \psi_{m+1}) 
\frac{z_{j:m}}{z_{m+1:N+j-1}}
\\ \nonumber
& + &
\omega^{-2}
\sum_{m=j+1}^{j+N-1} 
(\omega^{-2} \, \alpha_{m} \, \psi_{m} + \omega^2 \, \beta_{m+1} \, \psi_{m+1})
\frac{z_{j+1:m}}{z_{m+1:N+j}}
\end{eqnarray}
Since we will be concerned mainly with one permutation $\mathbf{p}$,
we introduce the more compact notation
$z_{i} \equiv z_{p_{i}}$ 
and 
$z_{i:k} \equiv \prod_{j=i}^{k} z_{p_{j}}$,
together with the cyclic condition
$z_{j} = z_{j+N}$.
Let us examine in details all the terms.
The first term in the rhs of (\ref{evolutionequation}) 
describes the process in which all the particles shift
one step to the right (Fig.~\ref{fig3}~a).
Notice that the special particle is the $j$th one in
the upper row and the $(j-1)$th in the lower row.
The weight of the process is $a$ because the arrow
crosses the seam in the right direction.
The product $z_{1:N}$ accounts for the
global shift to the right.
The second term describes the left translation 
(Fig.~\ref{fig3}~b).
Let us clarify the two sums. 
Start with the first one (Fig.~\ref{fig3}~c).
The special particle in the upper row is connected
via a clockwise half-loop to the left neighboring
ordinary particle.
This is taken into account by the coefficient 
$\omega^{2}$
in front of the sum.
The half-loop in the lower row must be inserted between
any two particles except the ones already connected 
by a half-loop in the upper row.
Therefore the sum runs from $m=j$ to $m=j+N-2$.
Moreover the particles distributed between the $j$th and
the $m$th particle undergo a shift to the right while
those positioned between $m+1$ and $j+N-1$ move one step
to the left.
These translations are taken into account by the products
$z_{j:m}/z_{m+1:N+j-1}$.
Since the half-loop in the lower row may turn clockwise or
anticlockwise,
each term in the sum splits in two parts:
$\omega^{-2} \, \psi_{m}$ 
for the anticlockwise half-loop and
$\omega^{2} \, \psi_{m+1}$
for the clockwise one.
The same argument applies to the second sum.
The coefficients $\alpha_{m}$ and $\beta_{m+1}$ ensure
that loops which wrap around the cylinder pick up the
correct weight.
They are set equal to 
$a$ ($a^{-1}$) 
when an arrow, of any of the two colors, in the intervening
row of horizontal edges, crosses the seam in the right (left) direction.
Their values depend on the relative position of the special
particle in the upper row, the seam, and the half-loop in the
lower row.
Following Fig.~\ref{fig3}~c we see that
when the lower half-loop is placed between
the special particle $j$ and the seam, the intervening arrow,
cutted by the seam, points to the left, so that
$\alpha_{m}=\beta_{m+1}=a^{-1}$.
When it is positioned on the right of the seam
$\alpha_{m}=\beta_{m+1}=a$.
In the special case in which 
the seam cuts the lower half-loop 
$\alpha_{m}=a^{-1}$ and $\beta_{m+1}=a$.

Our aim is the computation of the coefficient $\psi_{j}$
and the eigenvalue $\Lambda$.
Rearranging the terms, equation (\ref{evolutionequation})
may be expressed in a more compact form as:
\be
\label{eq1}
\Lambda \, \psi_{j}
=
- a \, z_{1:N} \, \omega^{4} \, \psi_{j} 
- \frac{1}{a \, z_{1:N} \, \omega^{4}} \, \psi_{j}
+
\left( z_{j}^{-1} \, \omega^{-2} +z_{j} \, \omega^{2} \right)
\, Q_{j}
\ee
where the quantity $Q_{j}$ is defined as:
\be
\label{eq2}
Q_{j} 
\equiv 
\sum_{m=j}^{j+N-1} 
(\omega^{-2} \, \alpha_{m} \, \psi_{m} + \omega^{2} \, \beta_{m+1} \, \psi_{m+1}) 
\frac{z_{j+1:m}}{z_{m+1:N+j-1}}
\ee
The term $Q_{j}$ generates all the oriented  black half-loops in the lower row,
that can be accommodated between two consecutive particles.
An alternative definition of $Q_{j}$ comes from equation (\ref{eq1}):
\be
\label{eq3}
Q_{j} 
= 
\frac
{
\Lambda 
+ 
a \, z_{1:N} \, \omega^{4}  
+ 
a^{-1} \, z_{1:N}^{-1} \, \omega^{-4}
}
{
z_{j}^{-1} \, \omega^{-2} + z_{j} \, \omega^{2}
} 
\psi_{j}
\ee
From the original definition of $Q_{j}$ (\ref{eq2}) it follows that:
\be
\label{eq4}
z_{j+1} \, Q_{j+1} - z_{j}^{-1} \, Q_{j} 
=
(\omega^{-2} \, \psi_{j} + \omega^2 \, \psi_{j+1})
\left(
a \, z_{1:N} - \frac{1}{a \, z_{1:N}}
\right)
\ee
Substituting in the lhs of (\ref{eq4})
the alternative expression for $Q_{j}$ given by (\ref{eq3}):
\begin{eqnarray}
\label{eq5}
\nonumber
& &
\left(
\Lambda + a \, z_{1:N} \, \omega^4 + a^{-1} \, z_{1:N}^{-1} \, \omega^{-4}
\right)
\left(
\frac
{
z_{j+1}^2
}
{
\omega^{-2} + z_{j+1}^{2} \omega^{2}
} 
\, 
\psi_{j+1}
-
\frac
{
1
}
{
\omega^{-2} + z_{j}^2 \omega^{2}
} 
\, 
\psi_{j}
\right)
\\
& = &
(\omega^{-2} \, \psi_{j} + \omega^2 \, \psi_{j+1})
\left(
a \, z_{1:N} - \frac{1}{a \, z_{1:N}}
\right)
\end{eqnarray}
Eq.~(\ref{eq5}) can be read as a recursive relation for $\psi_{j}$.
It simplifies defining the eigenvalue via a new variable $\mu$:
\be
\label{eq6}
\Lambda 
\equiv
-
\left[
a \, z_{1:N} \, (\omega^{4}-\mu) 
+
\frac{1}{a \, z_{1:N}}(\omega^{-4}+\mu)
\right]
\ee 
Replacing the eigenvalue 
$\Lambda$ in Eq.~(\ref{eq5})
by the definition (\ref{eq6}) 
and dropping the factor which contains $a \, z_{1:N}$,
the recursive relation for
$\psi_{j}$ reads:
\be
\label{eq7}
\frac{(\mu-\omega^4) \, z_{j+1}^2 - 1}{\omega^{-2}+z_{j+1}^2 \omega^2} \, \psi_{j+1}
=
\frac{\mu + z_{j}^2 + \omega^{-4}}{\omega^{-2} + z_{j}^2 \omega^{2}} \, \psi_{j}
\ee
Notice that relation (\ref{eq7}) 
does not involve the parameter $a$.
To see how the parameter $a$ enters in the wave function 
we have to inspect the case in which there are no 
ordinary particles between the seam and the special particle.
Notice that in this special case the seam may cut the 
upper black half-loop.
We skip the calculation and give the final result
for the structure of the coefficient $\psi_{m}$:
\be
\label{eq8}
\psi_{m}
=
s
\,
\frac{\omega^{-2}+\omega^{2} \, z_{m}^2}{(\mu-\omega^4) \, z_{m}^2-1}
\prod_{k=1}^{m-1} \frac{\mu + \omega^{-4} +z_{k}^2}{(\mu-\omega^4) \, z_{k}^2-1}
\ee
where $s=1$ if the special particle is located
left of the seam, and $s=a^{-2}$ if it is 
positioned on the right side.
\begin{figure}[!t]
\centerline{\epsfxsize=13cm \epsfbox{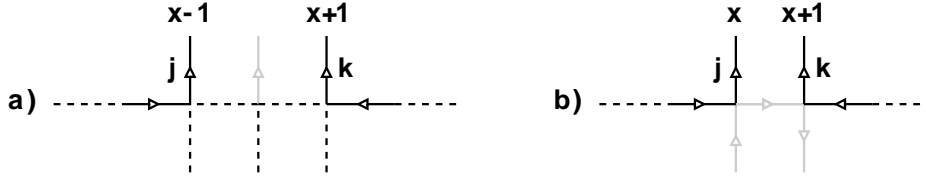}}
\caption{a) The two particles are at a distance of two, no black loop can be inserted in between.
b) Particles on adjacent sites force a gray half-loop to be inserted.} \label{fig4}
\end{figure}

The formalism developed up to now, 
works for sparsely distributed particles.
When the condition 
$x_{i+1}-x_{i}>2$ 
does not hold, several restriction on the amplitude 
$A_{\mathbf{p}}$ are necessary.
Suppose that two generic particles are located at position 
$x-1$ and $x+1$ (see Fig.~\ref{fig4}~a), 
then a half-loop of black arrows
can not be inserted between them in the lower row.
Thus, this event must be omitted from the generic list
of possibilities, in which the particles have both 
came from afar.
We demand the weight of the corresponding term, 
automatically generated by the lhs of the 
eigenvalue equation (\ref{CEE}), 
when the particles are sparse, 
to be zero in this special case:
\be
\label{eq10}
\sum_{j \leftrightarrow k} 
A_{\ldots j k \ldots} 
\left(
\omega^{-2} \, \alpha_{m} \, \psi_{m}
+ \omega^{2} \, \beta_{m+1} \, \psi_{m+1}
\right)
\,
z_{j}^x
\,
z_{k}^x
=
0
\ee
where the sum is over the interchange 
$j \leftrightarrow k$
of the indices and the shorthand notation
$A_{\ldots j k \ldots} \equiv A_{\ldots p_{j} p_{k} \ldots}$
has been used.
The way in which the momenta 
$z_{j}$ and $z_{k}$ 
enter in the coefficients 
$\psi_{m}$ and $\psi_{m+1}$
can be read from the structure of (\ref{eq8}).
Inserting (\ref{eq8}) in Eq.~(\ref{eq10})
and working out we see that only terms proportional to 
$\mu$ survive.
%
%
%
The resulting requirement for the scattering factor is:
\be
\label{eq11}
\frac{A_{\ldots j k \ldots}}{A_{\ldots k j \ldots}}
=
-
\frac
{
1+(\omega^4+\omega^{-4})z_{j}^2+z_{j}^2z_{k}^2
}
{
1+(\omega^4+\omega^{-4})z_{k}^2+z_{j}^2z_{k}^2
}
\ee
Let us make the following important observation:
the transfer matrix acts on a generic state
shifting each particle one step 
to the right or to the left.
Thus we can group particles in two sets
according to their position:
\emph{Odd} and \emph{Even} particles.
The two families do not mix.
This means that the scattering relation (\ref{eq11})
holds for particles belonging to the same family.
%
\begin{figure}[!t]
\centerline{\epsfxsize=15cm \epsfbox{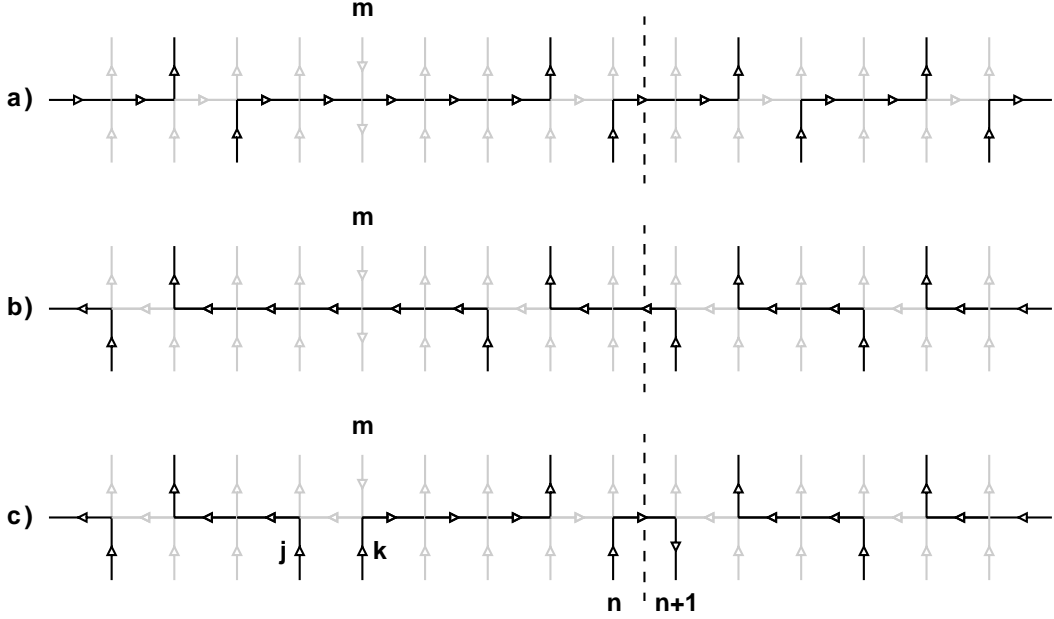}}
\caption{Four ordinary particles (black arrows up) 
and one bound state (gray arrow down).
The bound state is the second particle from the left ($m=2$).
The possible evolutions of the upper row are:
a) The bound state go straight and 
the ordinary particles shift to the right.
b) The bound state go straight and
the ordinary particles shift to the left.
c) The bound state forms a small gray half-loop
and evolves splitting into two black arrows (the $j$th and $k$th particles).
Periodic boundary conditions force a black-half loop
to be inserted in the lower row.
} \label{fig5}
\end{figure}

In the following we elaborate our formalism in order
to incorporate states containing grey arrows pointing down.
Start by noting that black arrows positioned on
adjacent sites generate a half-loop of grey arrows 
(see Fig.~\ref{fig4}~b).
This observation leads to treat the grey arrow down
as bound state of two generic particles.
Consider the states with $N-2$ ordinary particles
(black arrows up) and $1$ bound state (grey arrow down).
We make the following natural Ansatz on the form of the
wave function:
\be
\label{eq12}
\Phi_{m}(x_{1}, \ldots, x_{m}, \ldots x_{N-1})
\equiv
\sum_{\mathbf{p}}
A_{p_{1} \ldots p_{N}}
\phi_{m}(\mathbf{p})
\left[
z_{p_{1}}^{x_{1}}
\ldots
(z_{p_{m}} \, z_{p_{m+1}})^{x_{m}}
\ldots
z_{p_{N}}^{x_{N-1}}
\right]
\ee
where the index $m$ indicates that the bound state 
is preceded by $m-1$ ordinary particles.
The bound state is located at position $x_{m}$.
Notice that the wave functions 
(\ref{eq12}) and (\ref{BetheAnsatz}) have
essentially the same structure apart from the coefficient
$\phi_{m}$ which is still unknown at this stage.
In order to compute it we work out the eigenvalue equation
(\ref{CEE}) with the help of the ansatz 
(\ref{eq12}) and (\ref{BetheAnsatz})
for the particular case in which the initial state 
(rhs of Eq.~\ref{CEE})
contains two adjacent particles.
We see that if we require that:
\be
\label{eq13}
\sum_{j \leftrightarrow k} 
A_{\ldots j k \ldots}
(\omega^{-2} \psi_{m} + \omega^{2} \psi_{m+1}) z_{j} 
= 
\sum_{j \leftrightarrow k} 
A_{\ldots j k \ldots}
\phi_{m} (\eta^{-2} + \eta^{2} z_{j} z_{k})
\ee
then the generic list of possible evolutions 
(see rhs of Eq.~\ref{evolutionequation}) automatically
incorporates the gray half-loop.
The coefficient $\eta$ accounts for the two possible
orientations of the gray half-loop.
The next step is to study the evolution of the bound state.
The analog of 
Eq.~(\ref{evolutionequation}) 
for the bound state reads:
\begin{eqnarray}
\label{eq14}
\Lambda 
\,
\sum_{j \leftrightarrow k} 
A_{\ldots j k \ldots}
\phi_{m}
&  = &
\sum_{j \leftrightarrow k} 
A_{\ldots j k \ldots}
\left[
\phi_{m} 
\left( 
\frac
{
z_{j}z_{k}}{z_{1:N} \, a} + \frac{z_{1:N} \, a}{z_{j}z_{k}
} 
\right)
\right.
\\ \nonumber
& + &
\left.
\left(\eta^{-2} z_{k} + \frac{\eta^{2}}{z_{j}} \right)
\sum_{n=m+1}^{m+N-1} 
\frac{z_{m+2:n}}{z_{n+1:m+N-1}} 
(\omega^{-2} \, \alpha_{n} \, \psi_{n} + \omega^{2} \, \beta_{n+1} \, \psi_{n+1})
\right]
\end{eqnarray}
The bound state can move straight or 
form a half-loop as shown in 
Fig.~\ref{fig5}.
Now we have to check the consistency of 
Eq.~(\ref{eq14}).
For that eliminate $\phi_{m}$ in Eq.~(\ref{eq14})
using relation (\ref{eq13}).
In order to eliminate $\Lambda$, plug in the 
value of $\Lambda \, \psi_{j}$ given by the rhs of
the evolution equation (\ref{evolutionequation}).
At the end we get an expression relating $\psi_{j}$
for different values of $j$.
In the Appendix we will show that such expression is
consistent if we choose the same value for the
vertex weights
$\omega=\eta$ (in term of phases $e_{b}=e_{g}$)
and demand that the scattering amplitude
is symmetric for the interchange of particles 
belonging to \emph{different} families:
\be
\label{eq15}
A_{\ldots j k \ldots}
=
A_{\ldots k j \ldots}
\ee
Notice that the new scattering relation
(\ref{eq15}) 
is not in contradiction with the
previous relation
(\ref{eq11})
since the new one involves particles
of \emph{different} families.

We went further and studied the sector
with two special particles.
We skip the calculation and present the outcome.
Unwanted terms arise in the expression for the eigenvalue.
In order to cancel them we introduce another
amplitude $B_{\mathbf{q}}$, in the wave function, 
which regulates the behavior of the special particles.
We are forced to impose the following scattering relation:
\be
\label{eq16}
\frac{B_{\ldots i j \ldots}}{B_{\ldots j i \ldots}}
=
-
\frac{\mu_{i}\mu_{j}+\omega^{-4}\mu_{j}-\omega^{4}\mu_{i}}{\mu_{i}\mu_{j}+\omega^{-4}\mu_{i}-\omega^{4}\mu_{j}}
\ee
Now we have all the ingredients to derive 
the Bethe ansatz equations.
The final ansatz on the form of the wave function is:
\begin{eqnarray}
\label{eq17}
\Psi(x_{1},\ldots,x_{N}|r_{1},\ldots,r_{m})
& = &
\sum_{\mathbf{p},\mathbf{q}} A_{\mathbf{p}} B_{\mathbf{q}} 
\prod_{j=1}^{N} z_{p_{j}}^{x_{j}}
\\ \nonumber
& &
\prod_{i=1}^{m}
\left[
\frac{\omega^{-2}+\omega^{2}z_{p_{r_{i}}}^{2}}{(\mu_{q_{i}}-\omega^2)z_{p_{r_{i}}}^2-1}
\prod_{k=1}^{r_{i}-1} \frac{\mu_{q_{i}}+\omega^{-4}+z_{p_{k}}^2}{(\mu_{q_{i}}-\omega^{4})z_{p_{k}}^{2}-1}
\right]
\end{eqnarray}
The wave function has to satisfy periodic boundary conditions.
Suppose that the first particle is an ordinary one ($r_{1} \ne 1$):
\be
\label{PBC1}
\Psi(x_{1},\ldots,x_{N}|r_{1},\ldots,r_{m})
= 
\Psi(x_{2},\ldots,x_{N},x_{1}+L|r_{1}-1,\ldots,r_{m}-1)
\ee
From the Ansatz (\ref{eq17}), it follows that a sufficient
condition for the validity of (\ref{PBC1}), is that the complex
variable $z_{p_{1}}$ fulfills, for every permutation 
$\mathbf{p}$, the relation:
\be
\label{eq18}
A_{p_{1} \ldots p_{N}} 
\,
\prod_{i=1}^{m} 
\frac{\mu_{q_{i}}+\omega^{-4}+z_{p_{1}}^2}{(\mu_{q_{i}}-\omega^4)z_{p_{1}}^{2}-1}
=
A_{p_{2} \ldots p_{N} p_{1}}
\,
z_{p_{1}}^{L}
\ee
We make the additional assumption that the system 
has an even length $L$, this means that after 
having imposed periodic boundary conditions 
to a generic particle the sublattice to which the
particle belongs does not change.
We associate momenta $z_{j}$ ($j=1, \ldots , n_{z}$)
with the even sublattice
and $y_{j}$ ($j=1, \ldots , n_{y}$) with the odd sublattice.
By eliminating the amplitudes in (\ref{eq18}) using 
relations (\ref{eq11}) and (\ref{eq15}), 
we get the first two families of BA equations:
\be
\prod_{i=1}^{n_{\mu}} \frac{\mu_{i}+\omega^{-4}+z_{j}^2}{(\mu_{i}-\omega^4)z_{j}^{2}-1}
\,
\prod_{k \ne j}^{n_{z}}
-
\frac{1+(\omega^4+\omega^{-4})z_{j}^2+z_{j}^2z_{k}^2}{1+(\omega^4+\omega^{-4})z_{k}^2+z_{j}^2z_{k}^2}
=
z_{j}^{L}
\ee
\be
\prod_{i=1}^{n_{\mu}} \frac{\mu_{i}+\omega^{-4}+y_{j}^2}{(\mu_{i}-\omega^4)y_{j}^{2}-1}
\,
\prod_{k \ne j}^{n_{y}}
-
\frac{1+(\omega^4+\omega^{-4})y_{j}^2+y_{j}^2y_{k}^2}{1+(\omega^4+\omega^{-4})y_{k}^2+y_{j}^2y_{k}^2}
=
y_{j}^{L}
\ee
Similarly, assuming that the first particle is a special one,
we find an equation for the variable $\mu_{q_{1}}$:
\be
B_{q_{1} \ldots q_{m}}
=
B_{q_{2} \ldots q_{m} q_{1}}
\prod_{k=1}^{N}
\frac
{
\mu_{q_{1}}+\omega^{-4}+z_{p_{k}}^2
}
{
(\mu_{q_{1}}-\omega^4)z_{p_{k}}^{2}-1
}
\ee
Making use of relation (\ref{eq16}) we get
the third family of BA equations:
\be
a^{-2}
\,
\prod_{j=1}^{n_z} \frac{\mu_{l}+\omega^{-4}+z_{j}^2}{(\mu_{l}-\omega^{4})z_{j}^2-1}
\prod_{k=1}^{n_y} \frac{\mu_{l}+\omega^{-4}+y_{k}^2}{(\mu_{l}-\omega^{4})y_{k}^2-1}
\prod_{m \ne l}^{n_{\mu}} -\frac{\mu_{m} \mu_{l} + \omega^{-4} \mu_{l} - \omega^{4} \mu_{m}}{\mu_{m} \mu_{l} + \omega^{-4} \mu_{m} - \omega^{4} \mu_{l}}
=1
\ee
The factor $a^{-2}$ arises due to the fact that 
the special particle moves from the left to the right side
of the seam when periodic boundary conditions are imposed.

The eigenvalue expression that generalizes (\ref{eq6}) 
has the form:
\be
\Lambda
=
(-1)^{n_{\mu}}
\left[
a 
\, 
\prod_{j=1}^{n_{z}} z_{j} 
\prod_{k=1}^{n_{y}} y_{k} 
\prod_{m=1}^{n_{\mu}} (\omega^4-\mu_{m}) 
+
\frac{1}{a} 
\prod_{j=1}^{n_{z}} \frac{1}{z_{j}} 
\prod_{k=1}^{n_{y}} \frac{1}{y_{k}} 
\prod_{m=1}^{m_{\mu}} (\frac{1}{\omega^4}+\mu_{m}) 
\right]
\ee
\section{The free energy}
In order to find an exact solution of the model in the
thermodynamic limit we need difference kernels \cite{L1}.
First simplify a bit notation (\ref{relation1})
introducing the phase
$\theta \equiv e_{b}=e_{g}$ 
and reminding that
$a = \textrm{exp}(\textrm{i} \, \pi \, \alpha)$.
Then introduce a new set of variables $u_{j}, v_{j}, w_{j}$ related to $z_{j}, y_{j}, \mu_{j}$ by:
\be
z_{j}^{2}
=
\frac{\sin \frac{\pi \theta}{2}(1+u_{j} \, \textrm{i})}{\sin \frac{\pi \theta}{2}(1-u_{j} \, \textrm{i})}
\qquad
y_{j}^{2}
=
\frac{\sin \frac{\pi \theta}{2}(1+v_{j} \, \textrm{i})}{\sin \frac{\pi \theta}{2}(1-v_{j} \, \textrm{i})}
\qquad
\mu_{j}
=
\frac{2 \, \textrm{i} \, \sin \pi \, \theta}{1 - \textrm{exp}(\pi \, \theta \, w_{j})}
\ee
and define the function:
\be
\label{rational}
S_{c}(x,\theta)
\equiv
\frac
{
\sin\frac{\pi \, \theta}{2}(c+x \, \textrm{i})
}
{
\sin\frac{\pi \, \theta}{2}(c-x \, \textrm{i})
}
\ee
The BAE in the new variables are:
\be
S_{1}(u_{j},\theta)^{L/2}
=
-
\prod_{i=1}^{n_{w}} S_{1}(w_{i}-u_{j},\theta)
\prod_{k=1}^{n_{u}} -S_{2}(u_{j}-u_{k},\theta)
\ee
\be
S_{1}(v_{j},\theta)^{L/2}
=
-
\prod_{i=1}^{n_{w}} S_{1}(w_{i}-v_{j},\theta)
\prod_{k=1}^{n_{v}} -S_{2}(v_{j}-v_{k},\theta)
\ee
\be
\textrm{exp}(-2 \, \textrm{i} \, \pi \, \alpha)
\prod_{j=1}^{n_{u}} S_{1}(w_{l}-u_{j},\theta)
\prod_{k=1}^{n_{v}} S_{1}(w_{l}-v_{k},\theta)
\prod_{m=1}^{n_{w}} -S_{2}(w_{m}-w_{l},\theta)
=
-1
\ee
The eigenvalue is the sum of two terms:
\begin{eqnarray}
\Lambda
& = &
(-1)^{n_{w}}
\left[
\textrm{exp}(\textrm{i} \, \pi \, \alpha)
\prod_{j=1}^{n_{u}} S_{1}(u_{j},\theta)^{1/2}
\prod_{k=1}^{n_{v}} S_{1}(v_{k},\theta)^{1/2}
\prod_{m=1}^{n_{w}} 
\frac{\sin \frac{\pi \theta}{2}(w_{m} \, \textrm{i}-2)}{\sin \frac{\pi \theta}{2} w_{m} \, \textrm{i}}
\right.
\\ \nonumber
& + &
\left.
\textrm{exp}(-\textrm{i} \, \pi \, \alpha)
\prod_{j=1}^{n_{u}} S_{1}(u_{j},\theta)^{-1/2}
\prod_{k=1}^{n_{v}} S_{1}(v_{k},\theta)^{-1/2}
\prod_{m=1}^{n_{w}} 
\frac{\sin \frac{\pi \theta}{2}(w_{m} \, \textrm{i}+2)}{\sin \frac{\pi \theta}{2} w_{m} \, \textrm{i}}
\right]
\end{eqnarray}
Following the argument of \cite{L1} we see that the dominant eigenvalue
lies in the sector labeled by ($n_{u}=n_{v}=n_{w}=L/2$).
For $\alpha=0$ we get the largest eigenvalue when the roots are
symmetrically distributed on the real axis in such a way that no
holes appear between two consecutive roots (close packing of the roots).
Switching on the twist $\alpha$, the root distributions start to drift.
We verify numerically that:
\be
\prod_{j=1}^{n_{u}} S_{1}(u_{j},\theta)^{1/2}
\prod_{k=1}^{n_{v}} S_{1}(v_{k},\theta)^{1/2}
=
\textrm{exp}(\textrm{i} \, \pi \, \alpha)
\ee
Which allows for an eigenvalue expression with just one term:
\be
\label{eigenvalue}
\Lambda
=
A
\,
\prod_{m=1}^{n_{w}}
\left(
\cos^{2} \pi \, \theta + \sin^{2} \pi \, \theta \, \textrm{coth}^{2} \frac{\pi \, \theta \, w_{m}}{2}
\right)^{1/2}
\ee
And the $\alpha$ dependence has been encapsulated in the factor $A$:
\be
\label{factorA}
A 
\equiv
2 
\cos 
\left( 
2 \, \pi \, \alpha 
+ 
\sum_{m=1}^{n_{w}} 
\textrm{arctan} 
\left( 
\tan \pi \, \theta \, \textrm{coth} 
\left( 
\frac{\pi \, \theta \, w_{m}}{2} 
\right) 
\right)
\right) 
\ee
We can go further with the simplification taking the logarithm.
For that purpose define the function:
\be
\label{phic}
\phi_{c}(x,\theta) 
\equiv
\textrm{i} 
\, 
\log 
S_{c}(x,\theta)
\ee
And introduce the counting functions $Z_{L}(x,\theta)$ as follows:
\be
Z_{L}^{u}(u,\theta) \equiv -\frac{L}{2} \phi_{1}(u,\theta) + \sum_{m=1}^{n_{w}} \phi_{1}(w_{m}-u,\theta)
 - \sum_{k=1}^{n_{u}} \phi_{2}(u_{k}-u,\theta)
\ee
\be
Z_{L}^{v}(v,\theta) \equiv -\frac{L}{2} \phi_{1}(v,\theta) + \sum_{m=1}^{n_{w}} \phi_{1}(w_{m}-v,\theta) 
- \sum_{k=1}^{n_{v}} \phi_{2}(v_{k}-v,\theta)
\ee
\be
Z_{L}^{w}(w,\theta) \equiv  \sum_{j=1}^{n_u} \phi_{1}(u_j-w,\theta) + \sum_{k=1}^{n_{v}} \phi_{1}(v_{k}-w,\theta)
- \sum_{m=1}^{n_{w}} \phi_{2}(w_{m}-w,\theta)
\ee
Notice that the counting functions defined in this way have positive derivatives.
The next step is to define the root density function via 
the derivative of the counting functions in the thermodynamic limit:
\be
\rho(x) = \frac{1}{2 \, \pi} \,  \lim_{L \rightarrow \infty} \frac{d}{d x} \frac{Z_{L}(x)}{L}
\ee
Numerically we find that approaching the thermodynamic limit 
the roots spread over the real axis, from $-\infty$ to $+\infty$,
thus the integral equations for the root density functions are:
\be
\rho_{u}(u,\theta) 
=
-\frac{1}{4 \pi} \phi_{1}^{\prime}(u,\theta)
+ \frac{1}{2 \pi} 
\int_{-\infty}^{\infty} 
\left[
\phi_{1}^{\prime}(x-u,\theta) \rho_{w}(x, \theta)
-
\phi_{2}^{\prime}(x-u,\theta) \rho_{u}(x, \theta)
\right]
\,
dx 
\ee
\be
\rho_{v}(v,\theta) 
=
-\frac{1}{4 \pi} \phi_{1}^{\prime}(v,\theta)
+ \frac{1}{2 \pi} \int_{-\infty}^{\infty}
\left[
\phi_{1}^{\prime}(x-v,\theta) \rho_{w}(x, \theta)  
-
\phi_{2}^{\prime}(x-v,\theta) \rho_{v}(x, \theta)
\right]
\,
dx
\ee
\be
\rho_{w}(w,\theta) 
=
\frac{1}{2 \pi} \int_{-\infty}^{\infty}
\left[
\phi_{1}^{\prime}(x-w,\theta)
\, 
(
\rho_{u}(x, \theta)
+
\rho_{v}(x, \theta)
) 
-
\phi_{2}^{\prime}(x-w,\theta) \rho_{w}(x, \theta)
\right]
\,
dx
\ee
The system can be solved using Fourier transform.
The solutions for the root density functions are:
\be
\label{criticaldensity}
\rho_{w}(x)=\frac{1}{8} 
\, 
\textrm{sech}
\left(
\frac{\pi \, x}{4}
\right)
\qquad
\rho_{u}(x)=\rho_{v}(x)
=
\frac{\sqrt{2}}{8} 
\, 
\textrm{ch} 
\left(
\frac{\pi \, x}{4}
\right)
\,
\textrm{sech} 
\left(
\frac{\pi \, x}{2}
\right)
\ee
It is remarkable to notice that they 
do not depend on the parameter $\theta$.
There is still $\theta$ dependence in the expression 
for the largest eigenvalue (\ref{eigenvalue}).
The free energy in the thermodynamic limit has the following integral representation:
\be
\label{criticalfreeenergy}
f_{\infty}(\theta)
=
\frac{1}{16} \int_{-\infty}^{\infty}
\log 
\left[
\cos^2 \pi \, \theta + \sin^2 \pi \, \theta \, \coth^2 \frac{\pi \, \theta \, x}{2}
\right]
\,
\textrm{sech} \left( \frac{\pi \, x}{4} \right)
\,
dx
\ee
Notice that the $\alpha$ dependence (\ref{factorA}) drops out in the thermodynamic limit.
Applying Parseval formula we can drop the logarithm in the integrand and write
the free energy as a Laplace transform:
\be
\label{Laplace}
f_{\infty}(\theta)
=
-\int_{0}^{\infty}
\frac{e^{-p} \,\textrm{sech} (2 \, p \, \theta) \, \sinh (p \, \theta)}{p \, \sinh p}
\,
dp
-\frac{1}{2} \log 2 -\frac{3}{2} \log \pi +2 \log \Gamma(1/4)
\ee
For particular values of $\theta$ the integral can be solved $\cite{L1}$.
\newline
Two mutually excluding Hamiltonian walks \cite{JK1}:
\be
f_{\infty}(1/2)=\frac{1}{2} \, \log 2
\ee
The Ice model \cite{L1}:
\be
f_{\infty}(1/3)=\frac{3}{2} \, \log \frac{4}{3}
\ee
Four colorings on the square lattice \cite{KH,JK1}:
\be
\label{entropy4C}
f_{\infty}(\theta \rightarrow 0)=-\frac{1}{2} \log 2 -\frac{3}{2} \log \pi +2 \log \Gamma(1/4)
\ee
Notice that in this last particular limit 
expression (\ref{rational}) becomes rational.
These BA equations were derived for the first
time in \cite{BERNARD} and look very much like
those derived in \cite{M} for mixed \textrm{SU}(N)
vertex model for the case $N=4$.

An exact solution can be found also in the non critical region which corresponds to fugacity $n > 2$.
Since the fugacity is defined by $n=2 \, \cos \pi \, \theta$, this region corresponds to imaginary values
for the parameter $\theta$.
For convenience we assume $\theta$ real so that the fugacity
is now given by $n=2 \, \textrm{cosh} \, \pi \, \theta$, and redefine the function (\ref{rational}) and (\ref{phic}) by:
\be
\phi_{c}(x,\theta)
\equiv
\textrm{i} \, \log 
\left(
\frac{\textrm{sinh} \frac{\pi \theta}{2} (c+x \, \textrm{i})}{\textrm{sinh} \frac{\pi \theta}{2} (c-x \, \textrm{i})}
\right)
\ee
Now in the thermodynamic limit, the roots corresponding to the largest eigenvalue are symmetrically
distributed in the finite interval $[-1/\theta,1/\theta]$.
This time the integral equation can be solved by Fourier series.
The Fourier components are:
\be
\widehat{\rho_{u}}(m,\theta)=\widehat{\rho_{v}}(m,\theta)
=
\frac{1}{2} \, \cosh (m\, \pi \, \theta) \, \textrm{sech} (2 \, m \, \pi \, \theta)
\qquad
\widehat{\rho_{w}}(m,\theta)
=
\frac{1}{2} \, \textrm{sech} (2 \, m \, \pi \, \theta)
\ee
The root density functions:
\be
\label{NONcriticaldensity}
\rho_{w}(x,\theta)
=
\frac{\theta}{2} 
\, 
\sum_{m=-\infty}^{+\infty} \widehat{\rho_{w}}(m,\theta) 
\, 
\textrm{exp}( \textrm{i} \, m \, \pi \, \theta \, x)
\qquad
\rho_{u,v}(x,\theta)
=
\frac{\theta}{2} 
\, 
\sum_{m=-\infty}^{+\infty} \widehat{\rho_{u,v}}(m,\theta) 
\, 
\textrm{exp}(\textrm{i} \, m \, \pi \, \theta \, x)
\ee
These can also be written in terms of Jacobi elliptic functions:
\be
\rho_{w}(x,\theta)
=
\frac{\theta \, K(k)}{2 \, \pi} \, \textrm{dn} \, [x \, K(k) \, \theta, k]
\ee
\be
\rho_{u,v}(x,\theta)
=
\frac{\theta \, K(k)}{4 \, \pi}
\, 
\left(
\textrm{dn} \, [x \, K(k) \, \theta +\textrm{i} \, K(k) \, \theta,k]
+
\textrm{dn} \, [x \, K(k) \, \theta -\textrm{i} \, K(k) \, \theta,k]
\right)
\ee
where the modulus $k$ and the complete elliptic integral of
the first kind $K(k)$ are related to the phase $\theta$ by:
\be
\textrm{K}(k)
\equiv
\int_{0}^{\pi/2}[1-k^{2} \, \sin^{2} \phi]^{-1/2} \, d\phi
\qquad
\frac{K(\sqrt{1-k^{2}})}{K(k)}
=
2 \, \theta
\ee
The free energy:
\be
\label{NONcriticalfreeenergy}
f_{\infty}(\theta)
=
\frac{\pi \theta}{2}+
\sum_{m=1}^{\infty}\frac{e^{- m \,  \pi \, \theta} \, \textrm{sinh} (m \, \pi \, \theta)}
{m \, \textrm{cosh}(2 \, m \, \pi \, \theta)}
=
-\frac{1}{4} \, \log q
+
\sum_{m=1}^{\infty}
\frac{1}{m}
\frac{1-q^{m}}{q^{m}+q^{-m}}
\ee
where $q \equiv \textrm{exp}(-2 \, \pi \, \theta)$.
Taylor expanding the summand around $q=0$ and summing with
respect to $m$ we get the following product expansion for
the partition function:
\be
\label{PE}
\lim_{L, M\rightarrow \infty} Z^{1/L M}
=
e^{f_{\infty}(\theta)}
=
q^{-1/4}
\prod_{p=1}^{\infty}
\frac
{
(1-q^{2 \, (2 \, p - 1)})
\,
(1-q^{4 \, p - 1})
}
{
(1-q^{4 \, p - 3})
\,
(1-q^{4 \, p})
}
\ee
For $q \rightarrow 1$ we get the partition function for the
four colorings model:
\be
e^{f_{\infty}(0)}
=
\prod_{p=1}^{\infty}
\frac{(2 \, p -1)(4 \, p -1)}{2 \, p \, (4 \, p - 3)}
=
\frac{\Gamma^{2}(1/4)}{\sqrt{2} \, \pi^{3/2}}
\ee
in agreement with the expression (\ref{entropy4C}).

We have all the ingredients to discuss the singular behavior
of the free energy.
We can read the non analyticity of the partition
function from expression (\ref{PE}).
It appears that (\ref{PE}) 
has a solid wall of singularities on the unit circle \cite{B1}.
Following \cite{GLAS, DEGUCHI} we calculate the analytic continuation
of the critical free energy (\ref{Laplace}) into the non critical region
so that we can extract the singularity of the free energy near
the critical point $n=2$.
We find that near $n=2$ the free energy has an essential singularity:
\be
f_{sing} \propto \textrm{exp}
\left(
-\frac{\pi^2}{4 \, |n-2|^{1/2}}
\right)
\ee
This means that at the critical point $n=2$ the model undergoes
an infinite order phase transition of the Kosterlitz-Thouless type.
The same type of transition has been found in \cite{R} 
for a loop model on the hexagonal lattice.
Even if the free energy exhibits an essential singularity at $n=2$,
the right and left derivatives, of generic order, approach the same
values, in the limit $n \rightarrow 2^{\pm}$.
%

For the first derivative we get:
\be
\left.
\frac{\textrm{d} f_{\infty}(n)}{\textrm{d} n}
\right|_{n \rightarrow 2^{\pm}}
=
\frac{1}{24}
\ee
It has been shown that the first order derivative 
of the free energy with respect to the fugacity
is related to the average loop length $\overline{L}$ \cite{J1}.
We found $\overline{L}=12$, which is three times the
minimal loop length allowed by the square lattice,
in agreement with the conjecture made in \cite{J1}.
\section{Conclusions}
We have studied the bulk properties of the $\textrm{FPL}^{2}$
model in the case in which the two loop fugacities are the same,
both in the critical $(n \leq 2)$ and non critical $(n>2)$ region.
The model undergoes an infinite order phase transition,
of the Kosterlitz-Thouless type, at the critical point $n=2$.
The Ansatz we make on the wave function works only when
the vertex weights are the same $\omega=\eta$.
This is because the amplitude that describes the scattering
of particles belonging to different (odd and even) families,
does not factorize into two particles scattering amplitudes when $\omega \neq \eta$.
The details are in the Appendix.
It will be interesting to find the Yang-Baxter relation
on this solvable line and hopefully extend it to the whole
phase diagram.
On the solvable line the central charge 
and the scaling dimensions can be calculated
exactly using the non linear integral equation (NLIE) method.
\section{Appendix}Replacing $\phi_{m}$ in Eq.~(\ref{eq14}) by Eq.~(\ref{eq13}) 
the consistency requirement becomes:
\begin{eqnarray}
\label{ap1}
& &
\sum_{j \leftrightarrow k} 
A_{\ldots j k \ldots}
\left[
\left(
\frac{z_{j}  z_{k}}{z_{1:N} \, a} 
+ 
\frac{z_{1:N} \, a}{z_{j}  z_{k}} 
- 
\Lambda
\right)
(\omega^{-2} \, \psi_{m} + \omega^{2} \, \psi_{m+1}) \, z_{j} 
\right.
\\ \nonumber
&  &
+
\left.
(\eta^{-2} + \eta^{2} \, z_{j}  z_{k})
\left(
\eta^{-2} \, z_{k} + \frac{\eta^{2}}{z_{j}} 
\right)
\sum_{n=m+1}^{m+N-1} 
\frac{z_{m+2:n}}{z_{n+1:m+N-1}} 
(\omega^{-2} \, \alpha_{n} \, \psi_{n} + \omega^{2} \, \beta_{n+1} \, \psi_{n+1})
\right]
=
0
\end{eqnarray}
In order to eliminate 
$\Lambda$,
plug in  the value of 
$\Lambda \, \psi_{m}$
and
$\Lambda \, \psi_{m+1}$
given by the rhs of 
Eq.~(\ref{evolutionequation}):
\begin{eqnarray}
\label{ap2}
\nonumber
& &
\sum_{j \leftrightarrow k} A_{\ldots j k \ldots}
\left[
\left(
\frac{z_{j}z_{k}}{z_{1:N} \, a} + \frac{z_{1:N} \, a}{z_{j}z_{k}} 
+ z_{1:N} \, \omega^4 \, a + \frac{1}{z_{1:N} \, \omega^4 \, a}
\right)
( \omega^{-2} \psi_{m} + \omega^{2} \psi_{m+1} ) z_{j}
\right.
\\ 
& - &
z_{j} \,  (z_{j}^{-1} \omega^{-4} + z_{j}) 
\sum_{n=m}^{m+N-1}
(\omega^{-2} \, \alpha_{n} \, \psi_{n}+\omega^{2} \, \beta_{n+1} \, \psi_{n+1}) 
\frac{z_{m+1:n}}{z_{n+1:N+m-1}}
\\ \nonumber 
& - &
z_{j} \,  (z_{k}^{-1} + z_{k} \, \omega^{4}) 
\sum_{n=m+1}^{m+N} 
(\omega^{-2} \, \alpha_{n} \, \psi_{n} + \omega^{2} \, \beta_{n+1} \, \psi_{n+1}) 
\frac{z_{m+2:n}}{z_{n+1:N+m}}
\\ \nonumber
& + &
\left.
(\eta^{-2}  + \eta^{2} \, z_{j} z_{k})
\left(\eta^{-2}z_{k} + \frac{\eta^{2}}{z_{j}} \right)
\sum_{n=m+1}^{m+N-1} 
(\omega^{-2} \, \alpha_{n} \, \psi_{n} + \omega^{2} \, \beta_{n+1} \, \psi_{n+1})
\frac{z_{m+2:n}}{z_{n+1:m+N-1}} 
\right]
=
0
\end{eqnarray}
We look for a sufficient condition 
on the scattering amplitude $A_{\ldots j k \ldots}$
which ensures the validity of the equality (\ref{ap2}).
We start by considering the terms, in the sums,
that involve 
$\psi_{j}$ 
for 
$j \neq m , m+1$
and require their single contribution to be zero.
Working out we get the following condition:
\be
\label{ap4}
\sum_{j \leftrightarrow k} 
A_{\ldots j k \ldots}
\left[
\frac{1}{z_{j}} - \frac{1}{z_{k}} + z_{j} z_{k}^2 - z_{j}^2 z_{k} 
+ \left(  \eta^4 + \frac{1}{\eta^4} - \omega^4 - \frac{1}{\omega^4} \right) z_{k}
\right]
=0
\ee
When 
$\eta = \omega $
the expression in parenthesis is antisymmetric and
the sum vanishes if the scattering amplitude is symmetric
$A_{\ldots j k \ldots}=A_{\ldots k j \ldots}$.
Sorting out the terms $\psi_{m}$ and $\psi_{m+1}$ in
(\ref{ap2})
we got a more complicate expression which also goes
to zero on the previous conditions, but not for the
more general solution of (\ref{ap4}).



\begin{thebibliography}{200}
\bibitem{JK1}J. L. Jacobsen and J. Kondev,
                  \emph{Field theory of compact polymers on the square lattice.}
                  Nucl. Phys. \textbf{B532} (1998) 635. [cond-mat/9804048].

\bibitem{J1}J. L. Jacobsen, J. Vannimenus,
                  \emph{Finite average lengths in critical loop models.}
		  J.Phys. A \textbf{32} (1999) 5455. [cond-mat/9903242].

\bibitem{J2}J. L. Jacobsen, J. Kondev,
                  \emph{Conformal field theory of the Flory model of polymer melting.}
		   [cond-mat/0209247].

\bibitem{J3}J. L. Jacobsen, J. Kondev,
                  \emph{Transition from the compact to the dense phase of two-dimensional polymers.}
                  J. Stat. Phys. \textbf{96} (1999) 21. [cond-mat/9811085].

\bibitem{J4}J. L. Jacobsen,
                  \emph{On the universality of fully packed loop models.}
		  J. Phys. A \textbf{32} (1999) 5445. [cond-mat/9903132].

\bibitem{KH}J. Kondev and C. L. Henley, \emph{Four-coloring model on the square lattice:
                                        A critical ground state.}
                                        Phys. Rev \textbf{B}  Vol~\textbf{52}, (1995) 6628.

\bibitem{KONDEV1}J. Kondev and C. L. Henley, \emph{Kac-Moody symmetries of critical ground states.}
                                             Nucl. Phys. \textbf{B464} (1996) 540. [cond-mat/9511102].

\bibitem{KONDEV2}J. Kondev, J. de Gier, B. Nienhuis,
                  \emph{Operator spectrum and exact exponents of the fully packed loop model.}
		  J. Phys. A \textbf{29} (1996) 6489. [cond-mat/9603170].

\bibitem{KONDEV3}J. Kondev, J. L. Jacobsen,
                  \emph{Conformational entropy of compact polymers.}
		    Phys. Rev. Lett.  Vol \textbf{81}, No \textbf{14}
		   (1998) 2922.

\bibitem{KONDEV4}J. Kondev,
                  \emph{Liouville field theory of fluctuating loops.}
		    Phys. Rev. Lett. Vol \textbf{78},
		   (1997) 4320. [cond-mat/9703113].

\bibitem{BERNARD3}B. Nienhuis, 
                  \emph{Exact critical point and critical exponents of O(n) models in two dimensions.}
		    Phys. Rev. Lett.  Vol \textbf{49}, No \textbf{15}
		   (1982) 1062.

\bibitem{BERNARD}B. Nienhuis, \emph{Tiles and colors.}
		               J. Stat. Phys. \textbf{102} (2001) 981. [cond-mat/0005274].

\bibitem{L1}E. H. Lieb, \emph{Residual entropy of square ice.}
		   Phys. Rev. Vol~\textbf{162}, No~\textbf{1}
		  (1967) 162.

\bibitem{B1}R. J. Baxter, \emph{Colorings of a hexagonal lattice.}
		  J. Math. Phys. Vol~\textbf{11}, No~\textbf{3} (1970) 784.

\bibitem{KOSTOV}I. K. Kostov,
		  \emph{Exact solution of the three-color problem on a random lattice.}
 		   Phys. Lett. \textbf{B549},
		  (2002) 245. [hep-th/0005190].

\bibitem{B3}R. J. Baxter, \emph{q colourings of the triangular lattice.}
		  J. Phys. A \textbf{19} (1986) 2821.

\bibitem{B4}R. J. Baxter, \emph{Chromatic polynomials of large triangular lattices.}
		  J. Phys. A \textbf{20} (1987) 5241.

\bibitem{BA1}M. T. Batchelor, H. W. J. Bl\"ote,
		  \emph{Conformal anomaly and scaling dimensions of the O(n) model from
                        an exact solution on the honeycomb lattice.}
 		   Phys. Rev. Lett.  Vol \textbf{61}, No \textbf{2}
		  (1988) 138.

\bibitem{BA1bis}M. T. Batchelor, H. W. J. Bl\"ote,
		  \emph{Conformal invariance and critical behavior of the O(n) model on the honeycomb lattice.}
 		   Phys. Rev. \textbf{B} Vol \textbf{39},
		  (1989) 2391.

\bibitem{BA2}M. T. Batchelor, J. Suzuki and C. M. Yung,
		  \emph{Exact results for hamiltonian walks from the solution of the fully packed
                        loop model on the honeycomb lattice.}
 		   Phys. Rev. Lett.  Vol \textbf{73}, No \textbf{20}
		  (1994) 2646. [cond-mat/9408083].

\bibitem{BA3}M. T. Batchelor, B. Nienhuis, and S. O. Warnaar,
		  \emph{Bethe-ansatz results for a solvable O(n) model on the square lattice.}
 		   Phys. Rev. Lett.  Vol \textbf{62}, No \textbf{21}
		  (1989) 2425.

\bibitem{BA4}M. T. Batchelor, H. W. J. Bl\"ote, B. Nienhuis and C. M. Yung,
		  \emph{Critical behaviour of the fully packed loop model on the square lattice.}
 		   J. Phys. A \textbf{29} (1996) L399.

\bibitem{BLOTE1}H. W. J. Bl\"ote, B. Nienhuis,
		  \emph{Fully packed loop model on the honeycomb lattice.}
 		   Phys. Rev. Lett.  Vol \textbf{72}, No \textbf{9}
		  (1994) 1372.

\bibitem{BLOTE2}H. W. J. Bl\"ote, B. Nienhuis,
		  \emph{Critical behaviour and conformal anomaly of the O(n)
                   model on the square lattice.}
 		   J. Phys. A \textbf{22} (1989) 1415.

\bibitem{BLOTE3}H. W. J. Bl\"ote, M. P. Nightingale, B. Derrida,
		  \emph{Critical exponents of two-dimensional Potts and bond percolation models.}
 		   J. Phys. A  \textbf{14} (1981) L45.

\bibitem{R}N. Yu. Reshetikhin,
                  \emph{A new exactly solvable case of an O(n)-model
                        on a hexagonal lattice.}
		  J. Phys. A  \textbf{24} (1991) 2387.

\bibitem{M}M. J. Martins,
                  \emph{Integrable mixed vertex models from braid monoid algebra.}
                  In \emph{Satistical physics on the eve of the 21-st century}, eds.
                  M. T. Batchelor, L. T. Wille, Vol 14 of 
                  \emph{Series on advances in Statistical Mechanics}, World Scientific, Singapore 1999.
                   [solv-int/9903006].

\bibitem{D1}E. Domany, D. Mukamel, B. Nienhuis, A. Schwimmer,
                  \emph{Duality relations and equivalence for models with O(n) and cubic symmetry.}
                  Nucl. Phys. \textbf{B190} (1981) 279.

\bibitem{DOTSENKO}V. S. Dotsenko, J. L. Jacobsen, M. Picco,
                  \emph{Classification of conformal field theories based on coulomb gases. Application to loop models.}
                  Nucl. Phys. \textbf{B618} (2001) 523. [hep-th/0105287].

\bibitem{RAGHAVAN}R. Raghavan, C. L. Henley, S. L. Arouh,
                  \emph{New two-color dimers models with critical ground state.}
                  J. Stat. Phys. \textbf{86} (1997) 517. [cond-mat/9606220].

\bibitem{DIFRANCESCO}P. Di Francesco, E. Guitter, J. L. Jacobsen, 
                  \emph{Exact meander asymptotics: A numerical check.}
                  Nucl. Phys. \textbf{B580} (2000) 757. [cond-mat/0003008].

\bibitem{HIGUCHI1}S. Higuchi,
                  \emph{Compact polymers on decorated square lattices.}
		  J. Phys. A \textbf{32} (1999) 3697. [cond-mat/9811426].

\bibitem{HIGUCHI2}S. Higuchi,
                  \emph{Loop model with generalized fugacity in three dimensions.}
		  J. Phys. A  \textbf{33} (2000) 1661. [cond-mat/9907335].
\bibitem{GLAS}M. L. Glasser, D. B. Abraham, E. H. Lieb, 
                  \emph{Analytic properties of the free energy for the ``Ice'' models.}
		  J. Math. Phys. Vol~\textbf{13}, No~\textbf{6} (1972) 887.
\bibitem{DEGUCHI}T. Deguchi,
                 \emph{Introduction to solvable lattice models in statistical and mathematical physics.}
                  [cond-mat/0304309]
\end{thebibliography}
\end{document}